\documentclass[a4paper,11pt]{article}

\usepackage{amssymb}

\usepackage{fullpage}
\usepackage{hyperref}
\usepackage{authblk}
\usepackage{graphicx,xcolor,verbatim}
\usepackage{amssymb,amsfonts,amsmath}
\usepackage{dblfloatfix}
\usepackage{latexsym}

\newcommand*\xbar[1]{%
  \hbox{%
    \vbox{%
      \hrule height 0.5pt 
      \kern0.5ex
      \hbox{%
        \kern-0.1em
        \ensuremath{#1}%
        \kern-0.1em
    }%
    }%
  }%
}

\begin{document}
\title{Crystallization Inhibitors: Explaining Experimental Data through Mathematical Models}
\author[1]{M.P.~Bracciale}
\author[2]{G.~Bretti}
\author[1]{A.~Broggi}
\author[2]{M.~Ceseri}
\author[3]{A.~Marrocchi}
\author[2]{R.~Natalini}
\author[3]{C.~Russo}
\affil[1]{Dipartimento di Ingegneria Chimica Materiali Ambiente, Sapienza Universit\`{a} di Roma, Via Eudossiana 18, 00184, Rome, Italy}
\affil[2]{Istituto per le Applicazioni del Calcolo "Mauro Picone", via dei Taurini 19, 00185, Roma, Italy}
\affil[3]{Dipartimento di Chimica, Biologia e Biotecnologie, Universit\`{a} degli Studi di Perugia, Via Elce di Sotto 8, 06123, Perugia, Italy}

\maketitle

\begin{abstract}
In this paper we propose a new  mathematical model describing the effect of phosphocitrate (PC) on sodium sulphate crystallization inside bricks. This model describes salt and water transport, and crystal formation in a  one dimensional symmetry. This is the first study that takes into account mathematically the effects of inhibitors inside a porous stone. To this aim, we introduce two model parameters: the crystallization rate, which depends on the nucleation rate, and  the specific volume of precipitated salt. These two parameters are determined by numerical calibration of our system model for both the treated and non treated case.
\end{abstract}


\section{Introduction}

It is well known that one of the major causes of building degradation is the crystallization of salts into the porous matrix \cite{Goudie1997,Charola2000,Doehne2002,Schiro2012}. Salt is present inside building stones as free ions: it can be a natural element of the material, created by reaction with atmospheric pollutants or introduced by water solutions penetrating into the porous matrix by capillarity \cite{scherer2004}. The latter is the main mechanism leading to buildig damage and has received much attention from the scientific investigation \cite{RodriguezNavarro1999,Scherer2001,Flatt2002,EspinosaMarzal2010a,EspinosaMarzal2010b} but  remains not yet fully understood. 
Salt decay requires the simultaneous presence of soluble salts and water in the porous material, as well as appropriate environmental conditions. Indeed, it originates from salt-ions (e.g. chloride, nitrate, sulphate) that migrate while dissolved in liquid water which flows in the pore network of building materials. Liquid water may penetrate these materials by different processes, including hygroscopic moisture, penetration of rainwater (through, e.g. construction  joints, damaged roofs, and cracks), dew point condensation, and rising damp. The latter is probably the most frequent and perhaps one of the most difficult  sources of water to remove, when dealing with old buildings. Consider an initially dry porous stone (such as a masonry brick) that is wetted by a salt water solution. During the wetting phase, water fills up the stone bringing the dissolved salt present in the outside environment. If the stone is in contact with ambient air, water molecules are exchanged with the environment by evaporation thus starting a drying phase; the rate of dehydration depends on the relative humidity of the atmosphere. At this point, salt content in water increases and solution may become supersaturated. Once a high degree of supersaturation is reached, salt starts crystallizing: if crystals are formed inside the porous matrix we talk of \textsl{subflorescence} or \textsl{cryptoflorescence}; if crystallization takes place on the exterior boundaries of the stone we talk of \textsl{efflorescence}. \textsl{Subflorescence} causes the formation of large crystals into the pores: once the pres1e exerted by these crystals exceeds the tensile strength of the porous matrix, it can lead to widespread loss of surface, e.g. exfoliation, detachments. 
The occurrence of \textsl{efflorescence} or \textsl{subflorescence} (\textsl{cryptoflorescence}) depends on several factors including salt type and concentration, microclimate, evaporation rate \cite{RodriguezNavarro1999}, substrate porosity characteristics \cite{Rothert2007,Cultrone2008,Espinosa20081350,ShahidzadehBonn2010} and surface tension and viscosity of the solution \cite {RuizAgudo2007,Cardell2008,Sawdy2008}. 
The in-pore crystallization causes a reduction of the pore volume, breaking the liquid network and delaying water transport. Since pore clogging affects the location and quantity of crystals, it might have implications for stress development and deterioration of the material \cite{EspinosaMarzal2013}.
Common constructions contain different kind of salts such as chlorides, sulphates, nitrates, and carbonates, with their own solubility, crystalline structure and crystallization properties. Among these, sodium sulphate is probably one of the most complex and damaging salt types involved in salt decay processes. Indeed, this salt has three different phases of crystallization at various microclimate conditions, can easily supersaturate and has a solubility which is highly temperature dependent \cite{RodriguezNavarro2000,Steiger2008}. Both crystallization and hydration transformations in sodium sulphate, resulting in significant volumetric changes, have been blamed for the destructive mode of action of this salt \cite{Doehne2010}.

One way to prevent the stone breakage is to treat the porous material with a substance that inhibit \textsl{subflorescence}: these crystallization inhibitors reduce the pressure associated with the growing crystals trying to keep it below the breakage modulus of the substrate. The organic as well as inorganic ion and molecule additives alter the surface properties of the crystals which lead to changes in nucleation, growth, and thereby changes in the shape of the crystals as well as in their agglomeration/dispersion behaviour. Examples of well-known additives with extended technological and industrial uses  are the families of (poly)phosphates, carboxylates, polyacrylic acid derivatives, and benzotriazoles \cite{Amjad2001,Hasson2011,Kofina2007}. These additives are widely used as scale-inhibitors to prevent undesired effects associated with sparingly soluble salts (e.g. sulphates, carbonates) precipitating in oil extraction pipelines \cite{Black1991} industrial boilers, heat exchangers, house appliances or water pipes \cite{Garcia2001,Zafiropoulou2000} and others. 
The effectiveness of a given inhibitor depends on many variables: salt type, pore structure properties of the substrate, application methodology, the composition of the inhibiting solution to cite a few. Hence, a given modifier has to be evaluated for each stone and for each salt \cite{Rivas2010}. On the other hand, adding a crystal inhibitor does not affect surface tension nor contact angle of the wetting liquid, since there have not been observed any significant effect on solution transport \cite{RodriguezNavarro2002}.
Although the effectiveness of some salt crystallization inhibitors in bulk solution has been proved, the possibility of using these products for the prevention of salt decay in building materials is still controversial because it is not clear how these inhibitors act. However, experiments suggest two possible mechanisms \cite{Amjad1995,Amjad2001,vanderLeeden1995,Öner1998}: \textit{nucleation delay} enhances salt transport toward the surface, thus increasing efflorescence; \textit{crystal habit modification} by absorption on specific faces of a growing crystal that decreases crystal growth rate. 
Another matter of discussion is the fact that crystal reduction would result in higher supersaturated solutions. It has been speculated, but not actually observed, that in this case the inhibitor may promote salt precipitation at higher supersaturation levels and, hence, the quick formation of large crystals. Therefore, a modifier would eventually increase the crystal pressure and the risk of damage instead of reducing it \cite{lubelli2007}.
Our group has undertaken a broad research project \cite{Marrocchietal2006,Marrocchietal2007,Franceschinietal2013} focusing on the effects of environment-friendly, non-invasive inhibitor systems on saline solutions percolating and crystallizing in a porous media following evaporation, in order to develop a sound methodology suitable for addressing the conservation needs of different salt-weathered sites. Our attention has been particularly focused on the crystallization inhibition properties of functionalized polycarboxylates (i.e. maleate, citrate, phosphocitrate, tartrate), with an emphasis on the phosphorylated family members. Indeed we have demonstrated that phosphocitrate (PC) has been revealed to be one of the most promising inhibitors, because of its effectiveness in controlling the crystallization of different salts (i.e. sodium sulphate, sodium chloride, sodium nitrate, calcium carbonate) and salt mixtures in a wide range of porous materials and in various ambient conditions.   

In this work we developed a mathematical model describing the effect of phosphocitrate (PC) on sodium sulphate crystallization inside a brick's porous matrix. There are plenty of mathematical models describing salt crystallization in porous stone. They consists of 3D multiphase systems of equations for heat and mass transport with various degree of complexity. Some models might also couple the governing equations with other effects: osmosis, stress tensor deformations and latent heat release due to salt crystals formation  \cite{espinosa2008,koniorczyk2008,koniorczyk2013,castellazzi2013}.
For the present study we have developed a simple mathematical model of salt and water transport and crystal formation. In fact, we limit our research to the considerations on few available data, which can be obtained using simple laboratory equipments, and so  it would not have made sense to include further effects. Moreover, since the experiments were carried out in laboratory at constant temperature, we did not consider directly temperature variations; we just included evaporation rate into the porous stone simply by defining an appropriate sink term in the water balance equation. Actually, this work is a preliminary study to describe mathematically the effects of inhibitors inside a porous stone: to our knowledge, this is the first attempt to develop a mathematical model for the effects of crystallization modifiers. As we shall see, we identified two model parameters that will be crucial for the appropriate description of an inhibitor: 
\begin{description}
\item[$K_s$] the crystallization rate taking into account the nucleation rate; 
\item[$\gamma$] the specific volume of precipitated salt, describing the crystal habit modification.  
\end{description}
These two parameters will be determined by the numerical calibration of our model - i.e. by comparing our numerical results with the available experimental data - for both the treated and non treated case.
The remain of our paper is organized as follows: the second section will describe the materials considered and the experiments performed; the third and fourth sections will introduce the mathematical model
and describe the numerical scheme applied to solve the system equations; in section five we will describe our results. The paper ends up with few conclusions.

\section{Materials and Methods}
In this section, we will introduce the experimental settings we are going to consider \cite{Cassar2008}. Commercially produced brick is tested. Bulk density $\rho_{v}$ was determined by weighing and measuring of dimensions of dry prismatic samples. The matrix density $\rho_{mat}$ was measured by helium pycnometer. The porosity $n_0\left[\%\right]$ was calculated according to the equation
\begin{equation}\label{eqporosity}
	n_0 = 100 \cdot \left(1- \rho_v/\rho_{mat} \right).
\end{equation}
The porosity determined in this way is $28.51\%\pm 0.04\%$.

Pore size distribution was determined by mercury intrusion porosimetry (MIP) by Carlo Erba instrument on a about 1g of material. 
All experiments were performed in air conditioned laboratory at $25\pm2 ^\circ$C and $30\pm5\%$ RH.
Table \ref{tab:poredis} shows the pore size distribution of the considered brick.
\begin{table}[htbp]
\begin{center}
\begin{tabular}{c|c|c|c|c}

Pore Radius &&&&\\
Interval ($\mu$m)&0.001-0.01& 0.01-0.1& 0.1-1&1-10\\
\hline
Distribution (\%)&2.8&9.3&42.2&45.7\\

\end{tabular}
\caption{Pore size distribution in the brick under consideration.}
\label{tab:poredis}
\end{center}
\end{table}

\subsection{Experiment 1: brick's capillary absorption and drying test in pure water.}
This set of experiments were conducted, according to standard UNI EN $1925$ (Determination of water absorption coefficient by capillarity) and NORMAL $29/88$ (Drying Behaviour), without the presence of salt. It will serve as a control sample to test transport properties of the materials under study. 
The brick specimen has the form of a cube of side $5$ cm, is positioned in a bucket containing water and immersed for $3$ mm in height. 
The water absorption for capillarity, expressed in $g/cm^2$, is defined as the quantity of water absorbed by the specimen having the base surface in contact with water as a function of time $t$, with room temperature and pressure. At different time intervals the specimen is taken and tamponed only on the wet surface and then weighted until the variation in the quantity of absorbed water between two consecutive measurements, for a 24 hours interval, is less than $1\%$ of the water mass. 
The determination of the quantity of water absorbed by the specimen per time unit is given by $W=\frac{(m_i-m_0)}{S}$ expressed in $g/cm^2$, where $W$ is the quantity of water absorbed (expressed in $g$) and $S=25$ $cm^2$ is the surface of the specimen in contact with the porous frame.  
The experiment is applied to a number of specimen and then the average of the time dependent values $W$ obtained for the different specimens is computed. Finally we get the averaged quantity $Q(t_k)$, with $t_k$ the time instants expressed in $s^{1/2}$.

\subsection{Experiment 2:  brick's capillary absorption and drying test in a salt saturated water solution.}\label{sec:2.2}

\begin{figure}
\begin{center}
\includegraphics[width=5cm,height=7cm]{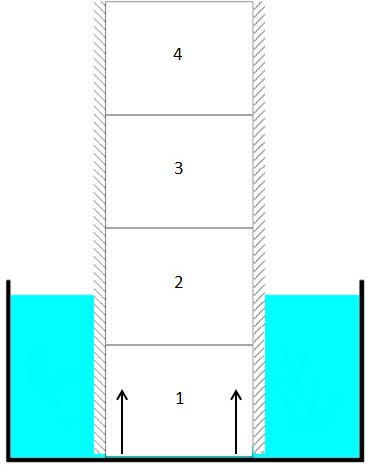}
\end{center}
\caption{Setup of experiment 2 as described in Section \ref{sec:2.2}.}
\label{fig:6}
\end{figure}

Both in untreated and treated brick's samples with PC the water and salt concentration profiles were determined experimentally using prismatic specimen $2\times 2.5\times 12$ cm positioned vertically in a bucket containing a salt water solution of $Na_2SO_4$ ($99.5$ g/L) (see Fig. \ref{fig:6}). In order to determine the concentration profiles the specimens were cut into 4 pieces with similar dimension and re-assembled sealing the lateral sides with epoxy resin; in this way only the top side of the brick is in contact with ambient air. On the other hand, the immersed part of the specimen is pervious and liquid can flow through the lateral side. The insulated specimens were dried at $65\pm2 ^\circ$C to the constant mass. When the solution in the bucket is totally absorbed by the specimens, the water content was obtained as difference of the mass of the saturated specimens and of the sample's mass after drying at $110\pm2 ^\circ$C to the constant weight. The concentration of sulphates in the dried samples was determined as follows: the samples were placed in plastic container, $200$ mL of boiling water was added and the container was sealed. This procedure was repeated every day for $1$ week. Then the dry samples were weighted and the concentration of sulphates was calcutated.

\section{The mathematical model}\label{sec:mathmodel}

Here we want to introduce a model of coupled water and sulphate transport taking into account not only the influence of water flow on salt transport but also the effect of bound sulphates on pore walls, and the effects of porosity changes (due to the salt bonding) on moisture transport. 
Regarding the mathematical domain, a reasonable assumption is to consider a one dimensional geometry since the domain is sealed on its lateral side; hence, flow is predominantly vertical.
We denote by $n$ the porosity, i.e. the fraction of volume occupied by voids, and we denote the fraction of volume occupied by the liquid and by the gas (composing the fluid) within the representative element of volume, respectively by $\theta_l$ and $\theta_g$. The following relation holds:
\begin{equation}\label{porosity}
n = \theta_l + \theta_g.
\end{equation}

 The mass balance equation for a liquid of density $\rho_l$ reads as:
\begin{equation}\label{waterbal}
\frac{\partial}{\partial t}(\rho_l\theta_l)+\frac{\partial}{\partial z}(\rho_l \mathbf{q})=f(\theta_l)
\end{equation} 
where $\mathbf{q}$ is the water flux into the porous matrix and $f(\theta_l)$ is the evaporation rate inside the specimen. Both $\mathbf{q}$ and $f(\theta_l)$ will be specified later on.


 Let us denote by $c_i$ the concentration of free ions in water and with $c_s$ the density of bound salt, the mass balance equation for salt dissolved in water is given by:
\begin{equation}\label{ionbal}
\partial_t(\theta_l c_i) +\partial_z (c_i\mathbf{q})=D\partial_z ( \theta_l \partial_z c_i) -\frac{\partial c_s}{\partial t},
\end{equation}
where $D$ is the salt diffusion coefficient, while the sink term on the right hand side takes into account the crystal formation into the porous matrix. 
In this work, we assume that crystal growth depends on the following properties: the concentration of salt dissolved in the liquid, the fraction $\theta_g$ and the degree of supersaturation.
If we indicate the supersaturation level with $\bar{c}$ we have:
\begin{equation}\label{crygrowth}
\frac{\partial c_s}{\partial t} = K_s c_i \theta_g^2 + \xbar K (c_i - \bar c)_+ \theta_l.
\end{equation}
with $K_s$ and $\xbar K$ two crystallization coefficient and $(\cdot)_+$ is the positive part function (or the second term is active only when salt saturation into the liquid exceeds the supersaturation level). The term $\theta_g$ on the right hand side is raised to power two in order to capture the following fact: the higher the water content, the smaller the crystallization into the pores. The power two simply slows down the crystal formation in saturated regions.  
The second term on the right hand side has been defined for the sake of completeness; in fact, in our experiments and in the subsequent simulations, salt supersaturation has never been exceeded and term $\xbar K$ has not been determined.

Since the overall porostity changes as the salts growth into the porous material, the following equation 
holds:
\begin{equation}
n(t) = n_0 - \gamma c_s,
\end{equation} 
with $\gamma$ the specific volume of sulphate crystal.

\subsection{Darcy's law}

Water flow into a porous medium is given by the well known Darcy's law \cite{barenblatt2010theory,bear1990introduction}:
\begin{equation}\label{Darcy} 
\mathbf{q}=-\frac{k(s)}{\mu_l} \left(\frac{n}{n_0}\right)^2 (\partial_z P_c(s) - \rho_l g)
\end{equation}
with $P_c=P_c(\theta_l/n)$ the capillary pressure, $k$ the permeability of the porous matrix, $\mu_l$ the viscosity of the fluid, the term $(n/n_0)^2$ is a shape factor for the influence of the porosity variation to the water flux and $s=\theta_l/n$.

\begin{figure}
\begin{center}
\includegraphics[width=9.5cm]{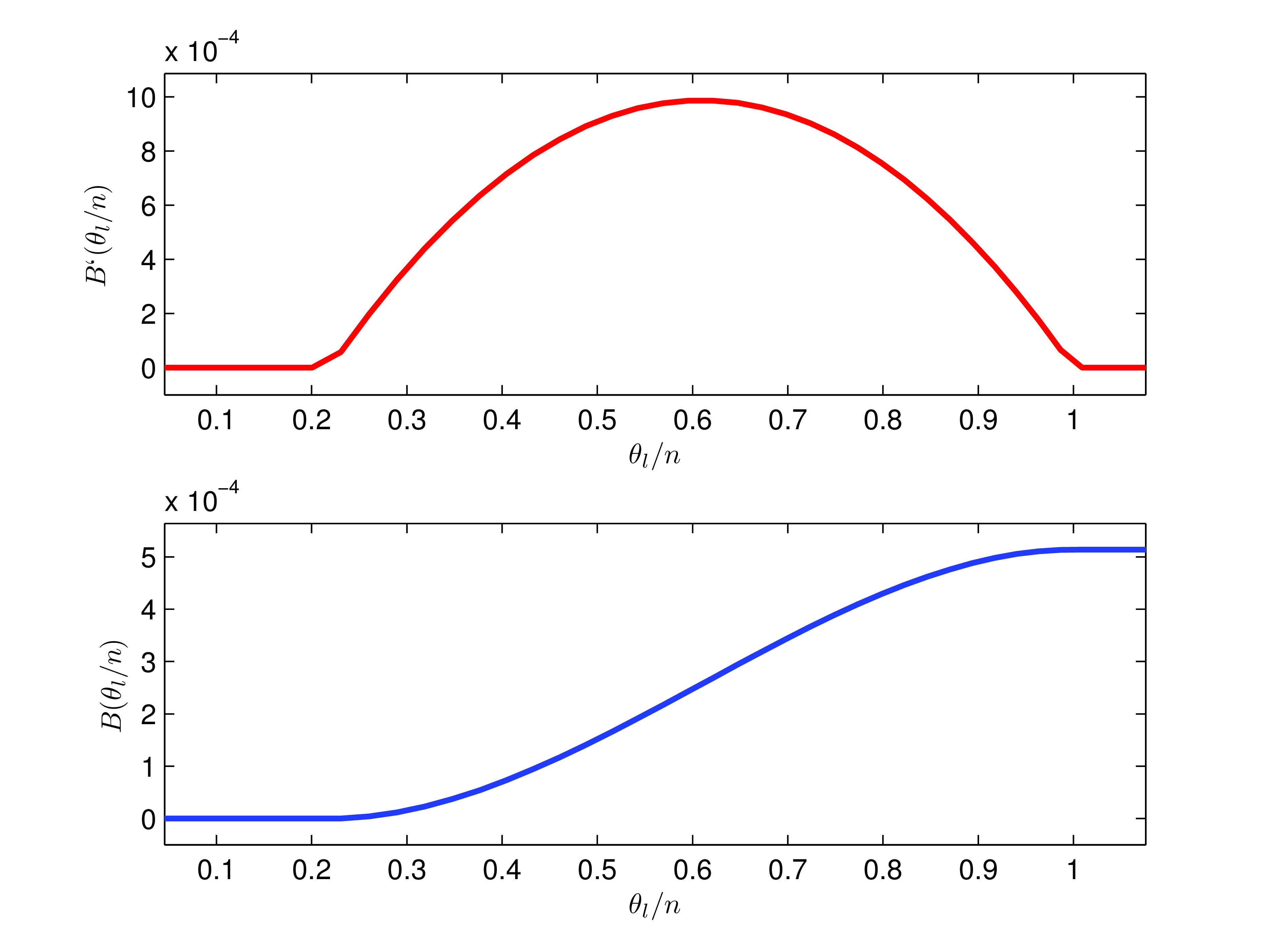}\\
\end{center}
\caption{Graph of the functions $B'_{a,c}(\theta)$ (above) and  $B_{a,c}(\theta)$ (below) for $s \in [a,1]$ for the choice $a=0.219$ and $c=9.87 \cdot 10^{-4}$.}
\label{fig:4}
\end{figure}

Capillary pressure is usually given as a function of water saturation and is defined through a state equation. 
In literature, one can find capillary pressure state functions for several applications; in building materials, however, despite the number of experimental study, there is not a relation correlating capillary pressure with moisture content into the porous matrix. To overcome this problem, we will approximate Darcy's law through a polynomial function with some free parameters that will be found through model calibration.   
Thus, we proceed as in \cite{Clarelli2010}.
First of all, since the dimensions of the brick are small, gravity effects can be safely disregarded from (\ref{Darcy}).
Then we introduce function $B$ such that
$$
\partial_z B=-\frac{k(\cdot)}{\mu_l}\partial_z P_c(\cdot).
$$
We know that $P_c(s)$ is a decreasing function of $s=\theta_l/n<1$ and vanishes whenever the medium is completely saturated, i.e. $\theta_l=n$. On the other hand, permeability $k=k(s)$ is a non-negative increasing function of $s$ and it is bounded from above by its value at saturation.
Taking into account these observations, the first derivative of function $B$ with respect to $s=\theta_l/n$ can be given by the ansatz
\begin{equation}
B'(s)=\max\left\{\frac{4c}{(1-a)^2}(a-s)(s-1),0\right\}
\end{equation}
with $a$ such that $k(a)=0$. Constants $a$ and $c$ are physical properties of the porous material involved and will be determined later on. The quantity $a\cdot n$ is the minimum value for saturation ensuring the hydraulic continuity - i.e. water transport through the porous medium. On the other hand $c$ has the dimensions of a diffusivity. The term $4c/(1-a)^2$ is chosen so that $\max\{B'(s)\}=c$.
Integrating $B'(s)$ we obtain the following expression (see Fig. \ref{fig:4}):

$$
B(s)= \left\{
\begin{array}{ll}
\frac{2}{3}c\left\{\left(\frac{1-s}{1-a}\right)^2(3a-1-2s)+(1-a)\right\},  \ \textrm{ if } s \in [a,1],\\
 0, \ \textrm{ if } s \in [0,a),\\
B(1)=\frac{2}{3}c(1-a), \ \textrm{ if } s>1.
\end{array}\right.
$$

Summing up, Darcy's law can be expressed as follows:
$$
\mathbf{q}=\left(\frac{n}{n_0}\right)^2\partial_z B\left(\frac{\theta_l}{n}\right)
$$

\subsection{Water evaporation}
Once water content decreases below the quantity $an$, the hydraulic continuity is broken and fluid trasport is no longer ensured. Since drying experiments end up with a completely dry stone, we added a sink term in the water balance equation (\ref{waterbal}) to take into account the effect of evaporation inside the porous matrix. In our mathematical model, we made the simplifying assumption that evaporation is maximum when moisture content is below the value $an$ and decreases quickly as the porous medium becomes saturated: thus liquid flow and evaporation acts at almost separated stages (one is strong while the other is weak and viceversa). This is reasonable since in our controlled experimental setting temperature is constant and does not play a significant role.   

\begin{figure} [htbp!]
\begin{center}
\includegraphics[width=8cm]{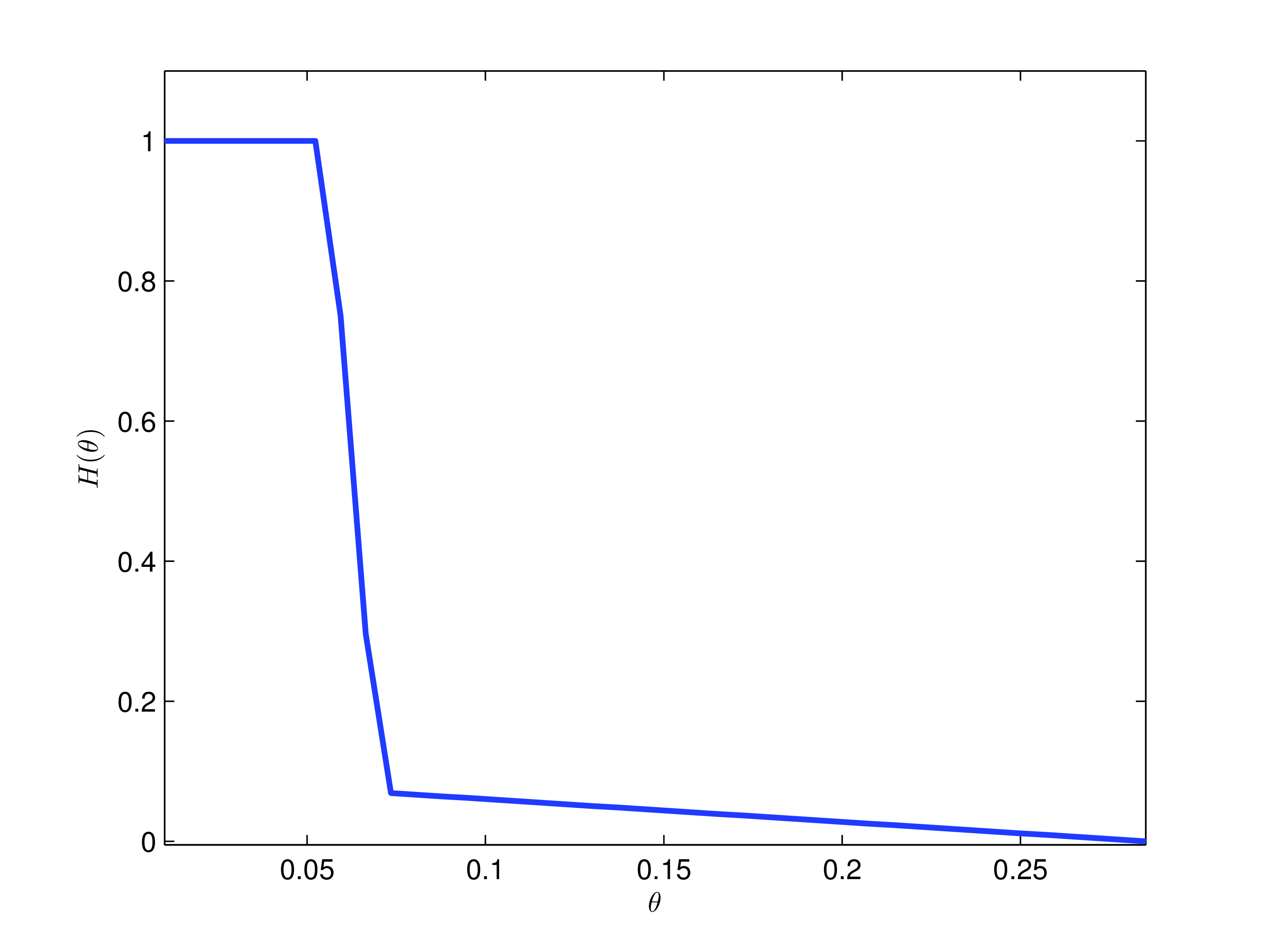}
\caption{Profile of the function $\mathcal H_\varepsilon(\theta)$, with $\varepsilon=0.25 \ a n$.}
\label{fig:funzH_new}
\end{center}
\end{figure}

We defined the evaporation rate as follows
\begin{equation}
f(\theta_l)=-\rho_l K_T\theta_l\mathcal{H}_\varepsilon(\theta_l)
\end{equation}
with $K_T$ a (temperature dependent) constant and $\mathcal{H}_\varepsilon$ is defined as follows:
\begin{equation} \label{funzH}
\mathcal H_\varepsilon(\theta)=\left\{
\begin{array}{ccc}
1& \text{ if } & 0< \theta < a n,\\
\frac{a n +\varepsilon-1}{\varepsilon} \theta + \frac{(an+\varepsilon)(1-an)}{\varepsilon}  & \text{ if } & an \le  \theta \le an+\varepsilon\\
 \frac{an+\varepsilon}{n(a-1)+\varepsilon} \theta - \frac{n(\varepsilon+an)}{n(a-1)+\varepsilon}  &\text{ if } & x>  an+\varepsilon .
\end{array}
\right.
\end{equation}
see Fig. \ref{fig:funzH_new}. In our simulations, we took $\varepsilon=0.25 \ a n$.

\subsection{The Complete Mathematical model} 
Summing up, the mathematical model we are going to consider is the following:
 \begin{equation}\label{modeleq}
 \left\{\begin{array}{ll}
\partial_t \theta_l = \partial_z \left(\left(\frac{n}{n_0}\right)^2 \partial_z B(\theta_l/n)\right)-K_T\theta_l\mathcal{H}_\varepsilon(\theta_l),\\
 \partial_t(\theta_l c_i) = \partial_z \left(c_i \left(\frac{n}{n_0}\right)^2 \partial_z B(\theta_l/n) + \theta_l D \partial_z c_i \right) - \frac{\partial c_s}{\partial t},\\ 
n(t) = n_0 - \gamma c_s,\\
\frac{\partial c_s}{\partial t} = K_s c_i (n-\theta_l)^2 + \xbar K (c_i - \bar c)_+ \theta_l.
 \end{array}\right.
\end{equation}

Table \ref{table:param1} shows the known parameters of the problem. Since some coefficients are unknown, we will calibrate the model versus experimental data. The obtained values will give some insight about the action of the inhibitor in the crystallization process. Table \ref{table:param2} lists the coefficients to be determined.

\begin{table*}[tbp]
\begin{center}
{
\begin{tabular}{|c|c|c|c|c|} \hline
  &  Description& Units & Value& Ref.\\\hline\hline
$h_1$  & Brick's height in the experiment 1& cm &  4.7&\\
$h_2$ & Brick's height immersed in the solution & cm &0.3&\\
$h_3$& Brick's height in the experiment 2&cm& 11.7&\\
 $n_0$ & Porosity of the unperturbed material & - &0.2851&Eq. \ref{porosity}\\
 $D$  & Diffusivity of Na$_2$SO$_4$& cm$^2$/s& $1.230\times 10^{-5}$&\cite[sect. 6.2]{jacob2002methods}\\ 
 $\rho_l$ & Density of water&g/cm$^3$&1&\cite{Chen2006}\\ 
$\bar\theta_l$ &Moisture content of the ambient air &g/cm$^3$&$6.254\times 10^{-2}$&Eq. \ref{thetabar}\\
$\bar c$&  Saturated concentration in water of sodium sulphate& g/cm$^3$&0.4399&Eq. \ref{crygrowth}\\ 
$\bar c_i$&  Concentration in water of sodium sulphate&g/cm$^3$&$9.95 \times 10^{-2}$&Eq. \ref{bottom1}\\ 
\hline\hline
\end{tabular} }
\vspace{0.2 in}    

\caption{Parameters of the problem.}\label{table:param1}
\end{center}
\end{table*}

\begin{table}[tbp]
\centering
{
\begin{tabular}{|p{1cm}|c|c|} \hline
  &  Description& Units \\
\hline\hline
$a$ & physical property of the porous matrix  & -\\
$c$ & physical property of the porous matrix  & cm$^2$/s\\
$\gamma$ &Specific volume of crystal& cm$^3$/g\\ 
$K_l$  & Exchange coefficient & cm/s \\ 
$K_s$& Crystallization rate coefficient &s$^{-1}$\\ 
$\xbar K$ & Growth rate of hydrated crystals & s$^{-1}$\\ 
$K_T$& Evaporation rate  & s$^{-1}$\\
$\alpha$ & Evaporation exponent & -\\
\hline\hline
\end{tabular} }
\vspace{0.2 in}    

\caption{Model coefficient to be calibrated}\label{table:param2}
\end{table}

\subsection{Boundary Conditions}
 For each experiment we will describe the initial and boundary conditions to apply to model (\ref{modeleq}). In some cases, we are even able to simplify the model equations.

\subsubsection{Experiment 1: pure water}
The immersed part of the brick (for $-h_2\leq z\leq 0$) is pervious to later water flow and we assume that it is initially saturated. In this way we can simply confine ourselves to mathematically describe the domain $0\leq z\leq h_1$.  Moreover, since there is no salt, our mathematical model reduces considerably; indeed, we can only retain the water continuity equation (\ref{waterbal}), that in this setting, is given by:
\begin{equation}\label{pb-water}
\partial_t \theta_l = \partial_{zz} B-K_T\theta_l\mathcal{H}_\varepsilon(\theta_l).
\end{equation}
Given the absence of salt, porosity will remain constant and, thus, will not affect water flow. Equation (\ref{pb-water}) has to be coupled with reasonable initial and boundary conditions.
For the experiment of imbibition, we assume the conditions 
\begin{equation}\label{imb-cond}
\left\{\begin{array}{ll}
\theta_l(z,0) = 0,\\
\theta_l(0,t)=n_0,\\
\end{array}\right.
\end{equation}
that is, the sample is initially dry while its botton side is always saturated.
To reproduce the loss of water at the upper boundary $z=h_1$ due to evaporation, we derive $\theta_l(h_1,t)$ from the following relations:
\begin{equation}\label{kk2}
\left\{\begin{array}{ll}
\partial_z B = K_l|\bar{\theta}_l - \theta_l|^{\alpha-1}(\bar{\theta}_l - \theta_l), \ \text{if } \theta_l > \bar\theta_l,\\
\theta_l = \bar{\theta}_l, \ \text{ otherwise}. 
\end{array}\right.
\end{equation}
In the above conditions, $\bar{\theta}_l$ is the moisture content of the ambient air (assumed constant) while $K_l$ is the exchange coefficient with the environment. The exponent $\alpha>1$ takes into account that water evaporation from the top of the doamin depends non-linearly on the difference between the quantity of water within the specimen and the value $\bar{\theta}_l$.
 
Once the imbibition stage is terminated, we stop the simulation and switch to another settings to deal with drying. In this case we consider the whole domain $[-h_2,h_1]$, since we do not add water at the bottom of the specimen. The other changes regard the initial and boundary conditions.
If we denote by $t_s$ the final time of imbibition and with $\theta_{fin}(z)=\theta_l(z,t_s)$ the value of $\theta_l$ after imbibition, the initial condition for the new setting is given by 
\begin{equation}\label{dry-initcond}
\left\{\begin{array}{ll}
\theta_l(z,0) = \theta_{fin}(z), \textrm{ for } z \in [0,h_1],\\
\theta_l(z,0) = n_0, \textrm{ for }z\in [-h_2,0]
\end{array}\right.
\end{equation}
meaning that the initial water content is the final value obtained for the imbibition test. Moreover, at  $z=-h_2$ we impose a no-flux boundary condition:
\begin{equation}\label{nullflux}
\partial_z\theta_l(-h_2,t) = 0,
\end{equation}
while at $z=h_1$ we retain condition (\ref{kk2}) again.


\subsubsection{Experiment 2: salt saturated water solution}
Experiments with salt solution were performed on bricks with height $12 \text{ cm}$. As above, we will consider during imbibition that the first three millimiters are submerged with water, thus we confine ourselves to the domain $[0, h_3]$, while during evaporation, to the domain $[-h_2,h_3]$ (see table \ref{table:param1}).
 
For imbibition, we assume the initial conditions for the system (\ref{modeleq}):
\begin{equation}\label{initcond}
\left\{\begin{array}{ll}
c_s(z,0) = 0,\\
c_i(z,0) = 0,\\
\theta_l(z,0) =\bar\theta_l,\\
n(z,0)=n_0.
\end{array}\right.
\end{equation}

As boundary conditions for $t \in[0,t_s]$,  we impose for the ion content, at $z=0$, the salt concentration $\bar c_i$ of the solution used in the experiment:
\begin{align}\label{bottom1}
c_i(0,t)&=\bar{c}_i
\end{align}

with $\bar{c}_i$ the actual concentration of sodium sulphate in water 
and a saturation condition for the water content
\begin{align}\label{bottom2}
\theta_l(0,t)&=n(0,t).
\end{align}

At the top boundary $z=h_3$, we impose:
\begin{align}\label{top1}
\partial_z c_i(h_3,t)&=0,
\end{align}

i.e. zero ion flux through the upper brick boundary and condition (\ref{kk2}).

For the drying phase, we assume the initial conditions:
\begin{equation}\label{initcond2}
\left\{\begin{array}{ll}
c_s(z,0) = c_s(z,t_s) ,\\
c_i(z,0) = c_i(z,t_s),\\
\theta_l(z,0) = \theta_l(z,t_s),\\
n(z,0)= n(z,t_s),
\end{array}\right.
\end{equation}
with $z \in [0,h_3]$ and for the immersed part, corresponding to $z \in [-h_2, 0]$, of the specimen we set:
 \begin{equation}\label{immersed}
\left\{\begin{array}{ll}
c_s(z,0) = 0,\\
c_i(z,0) = \bar c_i,\\
\theta_l(z,0) = n_0,\\
n(z,0)= n_0.
\end{array}\right.
\end{equation}
From now on we consider separately the four bricks composing the specimen. 
To this aim we define as $h^b_i$ the height of the broken brick and the points
$$
b_i= h^b_i,\text{ with } a_i=0\text{ for }i=1,2,3,4.
$$
The $i$-th brick is then parametrized as the interval $[a_i,b_{i}]\text{ for }i=1,2,3,4.$
Then as boundary conditions we impose at the bottom $z=a_i$, zero ion flux through the lower brick boundary
\begin{align}\label{bottom1dry}
\partial_z c_i(a_i,t)&=0, \ i=1,2,3,4
\end{align}
and as a boundary condition for ${\theta}_l$ reproducing the loss of water at the lower boundary we assume
\begin{equation}\label{theta_zero_bottom}
\theta_l(a_i,t)= \bar\theta_l, \ i=1,2,3,4.
\end{equation}
At the upper boundary $z=b_i$ we assume the conditions 
\begin{align}\label{up1dry}
\partial_z c_i(b_i,t)&=0, \ i=1,2,3,4
\end{align}
 and
\begin{equation}\label{theta_zero_up}
\theta_l(b_i,t)= \bar\theta_l, \, i=1,2,3,4
\end{equation}
and we put $\bar\theta_l=0$ in both conditions (\ref{theta_zero_bottom}) and (\ref{theta_zero_up}), in order to reproduce the situation inside the oven.

\subsubsection{Calculation of parameter $\bar{\theta}_l$}
Since we do not have any measurements of the relative humidity of the ambient air surrounding the sample, we set the value of the moisture content in the environment using the value of the average quantity of water within the brick measured in the imbibition-drying experiment with the sole water. In particular, using the measured average value $Q^s$ (quantity of water at saturation of the specimen) and the average value $Q^d=Q^{fin}-Q^s$ (loss of water at the end of the drying the experiment) we compute the final quantity of water $Q^{fin}=Q^s+Q^d= 0.31274$ $\ g/{cm}^2$ and then we get:
\begin{equation}\label{thetabar}
\bar\theta_l =\frac{Q^{fin}}{\rho_l h_1}=0.06254.
\end{equation}

\section{Numerical approximation}
Here we propose a numerical scheme for the model (\ref{modeleq}).
We mesh the interval $[0,h]$ with a step $\Delta z = \frac{h}{N}$ and we denote 
$$
\lambda = \frac{\Delta t}{\Delta z}, z_j = j \Delta z, j=1,...,N. 
$$
We also set $w^k_j=w(z_j,t_k)$ the approximation of the function $w$ at the height $z_j$ and at the time $t_k$. 
As showed in \cite{ADN} The simplest and consistent approximation of $\partial_z(r(z)\partial_z w)$ by means of Taylor expansions is the following first order approximation:
\begin{equation}
\begin{split}
\Delta_j(r,w) := \\
&\frac{(r_j + r_{j+1})(w_{j+1}-w_j) - (r_{j-1} + r_j)(w_j-w_{j-1})}{2 \Delta z^2}.
\end{split}
\end{equation}
From now on,  we will omit for simplicity the subscript $l$ of $\theta$. Then, the discretization in explicit form the first equation of the model (\ref{modeleq}) is:
\begin{equation}\label{eq1discr}
\frac{\theta^{k+1}_j- \theta^{k}_j}{\Delta t } =  \Delta_j ((n^k/n_0)^2, B^k)-\Delta t K_T \mathcal H_{\varepsilon}(\theta^k_j)\theta^{k+1}_j,\\
\end{equation}
Now, if we consider the velocity field computed in the equation (\ref{eq1discr}) and we set it as $V=(n/n_0)^2 \partial_z B(\theta/n) $, we can rewrite the second equation of the system (\ref{modeleq}) as:
\begin{equation}\label{eq2}
\partial_t(\theta c_i) - \partial_z (c_i V) = \partial_z (D \theta \partial_z c_i)  - K_s c_i (n-\theta) -\xbar K (c_i - \bar c)_+ \theta.
\end{equation}
We can assume:
\begin{equation}\label{Vdiscr}
 V^k_{j} = \left(\frac{n_j^k}{n_0}\right)^2 \frac{B\left(\frac{\theta^k_{j+1}}{n^k_{j+1}}\right)-B\left(\frac{\theta^k_{j-1}}{n^k_{j-1}}\right)}{2 \Delta z}, \ \textrm { for } j=1,\ldots,N-1,
\end{equation}
with the boundary values set as follows:
\begin{equation}\label{Vdiscr_bound1}
V^k_{0} = \left\{\begin{array}{ll}
0, \textrm{ for the imbibition phase},\\
 -\left(\frac{n_j^k}{n_0}\right)^2 K_l (\bar\theta_l-\theta^k_j), \textrm{ for the drying phase},
\end{array}\right.
\end{equation}
and
\begin{equation}\label{Vdiscr_bound2}
V^k_{N} = \left(\frac{n_j^k}{n_0}\right)^2 K_l (\bar\theta_l-\theta^k_j), \textrm{ for both phases.}
\end{equation}
Therefore, an explicit and monotonic scheme for (\ref{eq2}) reads as:
\begin{equation}
\begin{split}
\frac{(\theta c_i)^{k+1}_j- (\theta c_i)^{k}_j }{\Delta t } &= \frac{V^k_{j+1} c^k_{i,j+1}-V^k_{j-1} c^k_{i,j-1}}{2\Delta z}\\
& +\frac{|V^k_{j+1}| c^k_{i,j+1}- 2|V^k_{j}| c^k_{i,j} +|V^k_{j-1}| c_{i,j-1}^k}{2\Delta z}\\
&+\Delta_j(D\theta^k, c_i^k) - K_s c_{i,j}^k (n^k_j-\theta^k_j)\\
& - \xbar K (c_{i,j}^k - \bar c)_+ \ \theta^k_j,\\ 
\end{split}
\end{equation}
which is convergent under the CFL condition 
$$
\Delta t \le \frac{\inf \theta_j \Delta z^2}{D n_0 + sup |V| \Delta z  + (K_s + \bar K) n_0 \Delta z^2 }.
$$
We observe that the CFL may become very restrictive during the drying phase, since $\theta_j$ tends to zero. For this reason we simulated separately the two phases (imbibition and drying) using two different lower bounds for the CFL taking into account the evolution of $\theta_j$ in the two cases. 
Then, using the Euler's method for approximation of the third equation in (\ref{modeleq}),  we can write the discretized problem as:
\begin{equation}\label{modeleqdiscr}
\begin{split}
 \left\{\begin{array}{ll}
\theta^{k+1}_j &= \theta^{k}_j +  \Delta t \ \Delta_j ((n^k/n_0)^2, B^k)- \Delta t K_T \mathcal H_{\varepsilon}(\theta^k_j)\theta^{k+1}_j,\\ 
&\ j=1,\ldots,N-1\\
c_{s,j}^{k+1} &= c_{s,j}^k + \Delta t [K_s \ c_{i,j}^k (n^k_j-\theta^k_j) + \xbar K (c_{i,j}^k - \bar c)_+ \ \theta^k_j],\\  
& \ j=0,\ldots,N\\
n^{k+1}_j &= n_0 - \gamma c_{s,j}^{k+1}, \ j=0,\ldots,N\\
c_{i,j}^{k+1} &= \frac{1}{\theta^{k+1}_j}\left\{\theta^k_j c_{i,j}^{k} + \lambda \frac{|V^k_{j+1}| c_{i,j+1}^k- 2|V^k_{j}| c_{i,j}^k +|V^k_{j-1}| c_{i,j-1}^k}{2}  \right.\\ 
 &  \left.  + \Delta t \Delta_j(D\theta^k,c_i^k) + \lambda \frac{V^k_{j+1} c_{i,j+1}^k-V^k_{j-1} c_{i,j-1}^k}{2} -  \right.\\  
& \left. \Delta t [K_s c_{i,j}^k (n^k_j-\theta^k_j) + \xbar K (c_{i,j}^k - \bar c)_+ \ \theta^k_j]\right\},\\ 
 & \ j=1,\ldots,N-1,
 \end{array}\right.
\end{split}
\end{equation}

with suitable boundary conditions described in the next subsections. In particular, for the first equation of the scheme we have:
\begin{equation}\label{implth}
 \theta^{k+1}_j = C (\theta^{k}_j +  \Delta t \ \Delta_j ((n^k/n_0)^2, B^k)) 
\end{equation}
with 
$$
C=\frac{1}{1 + \Delta t K_T \mathcal H_{\varepsilon}(\theta^k_j)}.
$$

Note that the scheme in the last equation of (\ref{modeleqdiscr}) may become degenerate if $\theta^{k+1}_j$ is null, thus we put into the numerical algorithm a threshold for $\theta$ in order to avoid this possibility.

\subsection{Boundary conditions for the imbibition phase}
At the bottom boundary of the brick, we assume the condition for the ion content according to the concentration value of the experiment  (\ref{bottom1}), which reads as
\begin{align}\label{imb_ux0}
c_{i,0}^{k+1} &= \bar{c}_i
\end{align}
and the condition (\ref{imb_thetax0}).
At the top boundary of the brick, we impose the zero ion flux condition (\ref{top1}) for the ion content, discretized with a second order approximation:
\begin{align}\label{uxN}
c_{i,N}^{k+1}= \frac{4}{3} c_{i,N-1}^{k+1}- \frac{1}{3}c_{i,N-2}^{k+1}.
\end{align}
Let us now consider the discretization of the condition (\ref{kk2}), reproducing the exchange with the environment. Note that in the case of the experiment 1 with sole water in the condition (\ref{kk2}) we have to replace $n^k_j$ with the constant value $n_0$.\\
At the node $z_N$ we need to solve the equation
\begin{equation}
\begin{split}
\frac{3}{2 \Delta z} B(\theta^{k+1}_{N}/n^{k+1}_{N}) + K_l|\bar{\theta}_l - \theta^{k+1}_{N}|^{\alpha-1}(\bar{\theta}_l - \theta^{k+1}_{N}) =\\
 \frac{4 B(\theta^{k+1}_{N-1}/n^{k+1}_{N-1})-B(\theta^{k+1}_{N-2}/n^{k+1}_{N-2})}{2\Delta z},
\end{split}
\end{equation}
with the function to be inverted
$$
g_1(\theta) = \frac{3}{2 \Delta z} B(\theta/n) - K_l|\bar{\theta}_l - \theta|^{\alpha-1}(\bar{\theta_l} - \theta) .
$$
The invertibility condition is
\begin{equation} \label{invcond1}
 g'_1=\frac{3}{2 n \Delta z} \partial_{\theta} B(\theta/n) + K_l \alpha|\bar{\theta}_l - \theta|^{\alpha-1} > 0
\end{equation}
on a compact set, with 
\begin{equation}\label{Bx}
\begin{split}
\partial_{\theta} B(s=\theta/n) &=\\
&\left\{\begin{array}{ll} d\left(\frac{a+1}{n_0}\frac{\theta}{n_0} - \left(\frac{\theta}{n_0}\right)^2 \frac{1}{n_0} - \frac{a}{n_0} \right),  \ \textrm{ if } s \in [a,1],\\
 0, \ \textrm{ elsewhere }.
\end{array}\right. 
\end{split}
\end{equation}
Note that the condition $\partial_\theta B(\theta/n)>0$ is always satisfied for $\theta \in [a \cdot n,n]$, so that (\ref{invcond1}) holds.  Therefore, at the upper boundary of the brick we need to solve, using for example with Newton's method:
 \begin{equation}\label{newt2}
\theta^{k+1}_{N} = g_1^{-1}\left(\frac{4 B(\theta^{k+1}_{N-1}/n^{k+1}_{N-1})-B(\theta^{k+1}_{N-2}/n^{k+1}_{N-2})}{2\Delta z} \right).
\end{equation}

\subsection{Boundary conditions for the drying phase}
 
 In order to model the loss of water, we use the zero ion flux at the bottom of the brick, discretized with a second order approximation as
 \begin{equation}\label{dry_ux0}
 c_{i,0}^{k+1} = \frac{4}{3} c_{i,1}^{k+1} - \frac{1}{3}c_{i,2}^{k+1},
 \end{equation}
 and condition (\ref{uxN}) at the top boundary.
Let us now consider the discretization of the conditions (\ref{theta_zero_bottom}) and (\ref{theta_zero_up}), reproducing the situation of the specimen inside the oven we set at the lower boundary:
  \begin{equation}\label{dry_thetax0}
 \theta^{k+1}_{0} = \bar\theta_l,  
 \end{equation}

 and analogously at the upper boundary:
  \begin{equation}\label{dry_thetaxN}
 \theta^{k+1}_{N} = \bar\theta_l, 
 \end{equation}
with $\bar\theta_l=0$.


\section{Numerical Results and comparison with experimental data}
\subsection{Calibration of parameters $a, c, K_l, K_T, \alpha$.}
Now we describe the calibration procedure to determine $a$, $c$, $K_l$, $K_T$ and $\alpha$ for both the phases of imbibition and evaporation of water in the brick using the experimental data of experiment 1.\\ We need to compute the total quantity of water absorbed and lost by the brick at time $t_k$ given by:
\begin{equation}\label{int}
\int_{0}^{h_1} \rho_l \theta(z,t_k) dz,
\end{equation}
thus we need to solve problem (\ref{pb-water}). We compute $\theta(z,t_k)$ numerically with the forward-central approximation scheme
$$
\theta^{k+1}_j = \theta^k_j + \frac{\Delta t}{{\Delta z}^2} (B_{a,c}(\theta^k_{j+1}/n_0)-2B_{a,c}(\theta^k_j/n_0)+B_{a,c}(\theta^k_{j-1}/n_0))
$$
with the boundary condition at the top boundary (\ref{newt2}) under the CFL condition 
$$
\frac{\Delta t}{{\Delta z}^2} \le \frac{n_0}{2 \partial_z B_{a,c}}= \frac{n_0}{2 c },
$$
with $\theta^k_j =\theta(z_j, t_k)$, $z_j= j\Delta z, j=0,...,N=\left[\frac{h_1}{\Delta z}\right] $, $\{t_k\}_{k=1,...,N_{meas}}$. At the bottom boundary we use the imbibition condition
\begin{align}\label{imb_thetax0}
\theta^{k+1}_0 &= n^{k+1}_0
\end{align}
and Neumann condition $\theta_z(0,t)=0$ of null flux, only for the drying phase, that numerically results to be  
\begin{equation}
\theta^{k+1}_{0} = \frac{4}{3} \theta^{k+1}_1 - \frac{1}{3}\theta^{k+1}_{2}.
\end{equation}
Let us define $t_{s}$ the saturation time at the end of the imbibition phase and ${Q^s}$ the corresponding value.
 Then we compute the approximated values of the quantity of water in the brick  ${Q}^{num}_k$ at time $t_k$ as follows. With the trapezoidal rule we compute the integral (\ref{int}):
$$
{Q^{num}_k} = \rho\frac{\Delta z}{2} \left(\theta^k_0 + 2\sum_{j=1}^{N-1} \theta^k_j + \theta^k_{N} \right),
$$
in order to compare the numerical quantity of water to experimental data $Q_k$ at time $t_k$. 
The error to be minimized is then defined as
$$
E(a,c,K_l, K_T, \alpha) = \frac{1}{N_{meas}}\sum_{k=1}^{N_{meas}} \frac{|{Q^{num}_k} - Q_k|}{|\bar Q|},
$$
with $\bar Q$ the average value among data.
The calibration procedure has been carried out in MATLAB\textcircled{c} applying the simulated annealing method. The computational time for a single simulation with fixed parameters takes 900 seconds on an Intel(R) Core(TM) i7-3630QM CPU 2.4 GHz. Table \ref{table:2.1} lists the results obtained within an error of about $7\%$.

\begin{figure}
\begin{center}
\includegraphics[width=8cm]{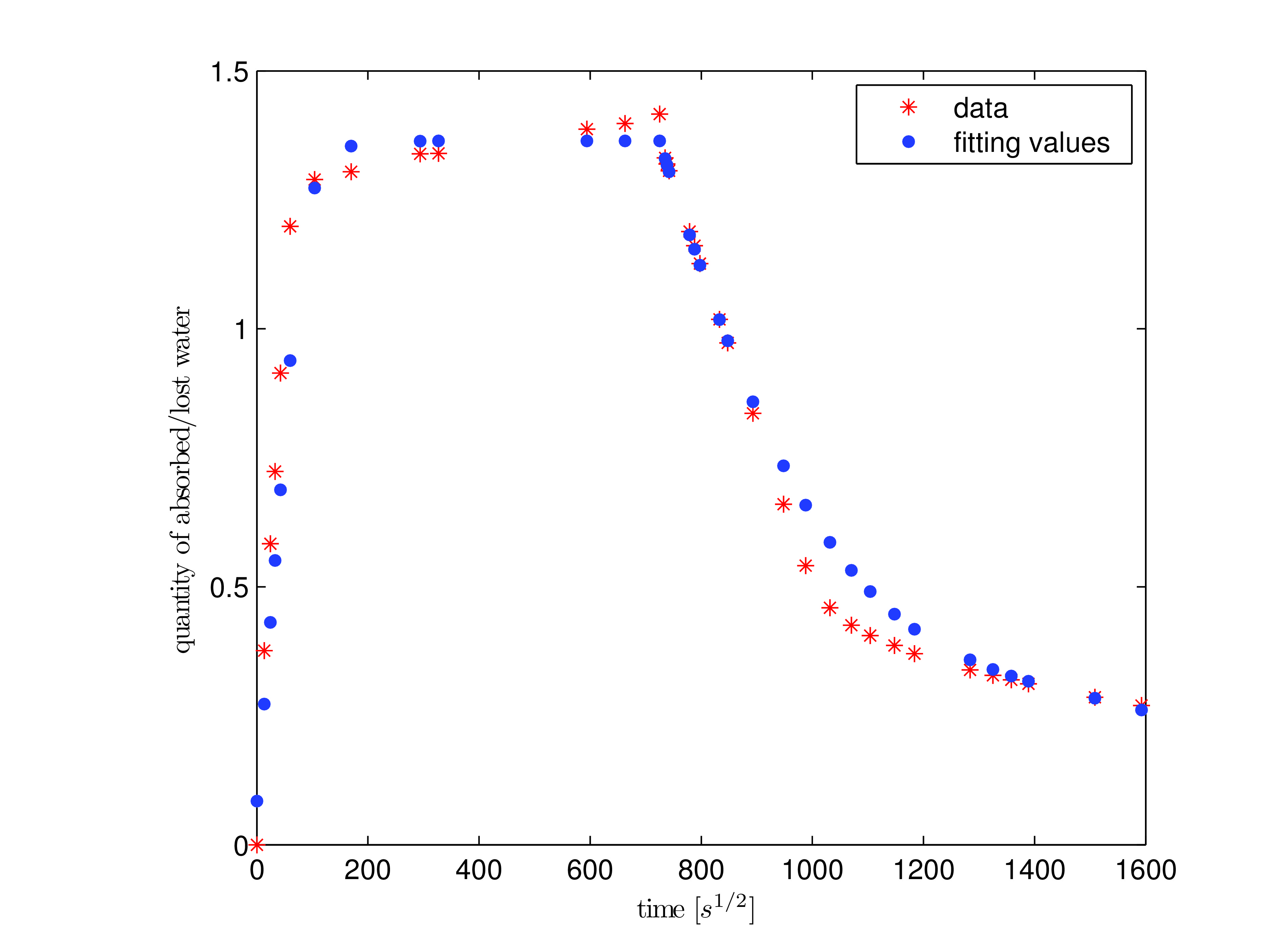}                                      
\end{center}
\caption{Data fitting result: comparison between data points and fitting values obtained for $a=0.21904, c=9.8073 \times 10^{-4}, K_l=10^{-5}, K_T= 3.2 \times 10^{-7}, \alpha=0.9$.}
\label{fig:5}
\end{figure}

\begin{table}
\begin{center}
\begin{tabular}{c|c|c}
Quantities & Value & Dimensions\\
\hline
$a$ & $0.21904$ & -\\
$c$ & $9.8073\times 10^{-4}$ & $\text{cm}^2 \text{s}^{-1}$\\
$K_l$ & $3\times10^{-5}$ & $\text{s}^{-1}$\\
$K_T$ & $3.2 \times 10^{-7}$ & $\text{s}^{-1}$\\
$\alpha$ & $0.9$ & -
\end{tabular}
\end{center}
\caption{Results of the calibration for the imbibition and drying stages without salts. The overall error is about $7\%$.}
\label{table:2.1}
\end{table} 
Figure \ref{fig:5} shows the comparison between measured data and numerical simulations after calibration.

\subsubsection{Calibration of constants $K_s$ and $\gamma$}
As described in Section \ref{sec:2.2} for experiment 2, the bricks were first broken in four pieces with similar dimensions, both for the treated and non treated cases; for any brick, we measured its salt content.

In order to determine constant $K_s$ and $\gamma$ we need to define an appropriate functional to be minimized. We proceed as follows.
First we define the average quantity of salt in $i$-th brick  as:
\begin{equation}\label{quant_salt}
\frac{A_i}{B_i}\int_{a_i}^{b_i}c_s(z,\bar{t})dz = \frac{1}{h^b_i}\int_{a_{i}}^{b_i}c_s(z,\bar{t}) dz\text{ for }i=1,2,3,4.
\end{equation}
where $\bar{t}$ is a sufficiently long time when we can assume that the water is completely evaporated.
Here $A_i$ and $B_i$ represent the cross section and the volume of brick $i$, respectively.\\
If we denote by $q^{num}_i$ the average quantity of salt in $i$-th brick obtained discretizing formula (\ref{quant_salt}) with the trapezoidal quadrature rule, the values of $K_s$ and $\gamma$ can then be found solving the following minimization problem
\begin{equation}\label{eq:functionalKs}
\min_{K_s,\gamma}\frac{1}{4}\sum_{i=1}^4\frac{|q^{num}_i-q_i|}{|\bar q|},
\end{equation}
with $\bar q$ the average salt content among the four bricks.

\begin{figure} [htbp!]
\begin{center}
\includegraphics[width=8cm,height=8cm]{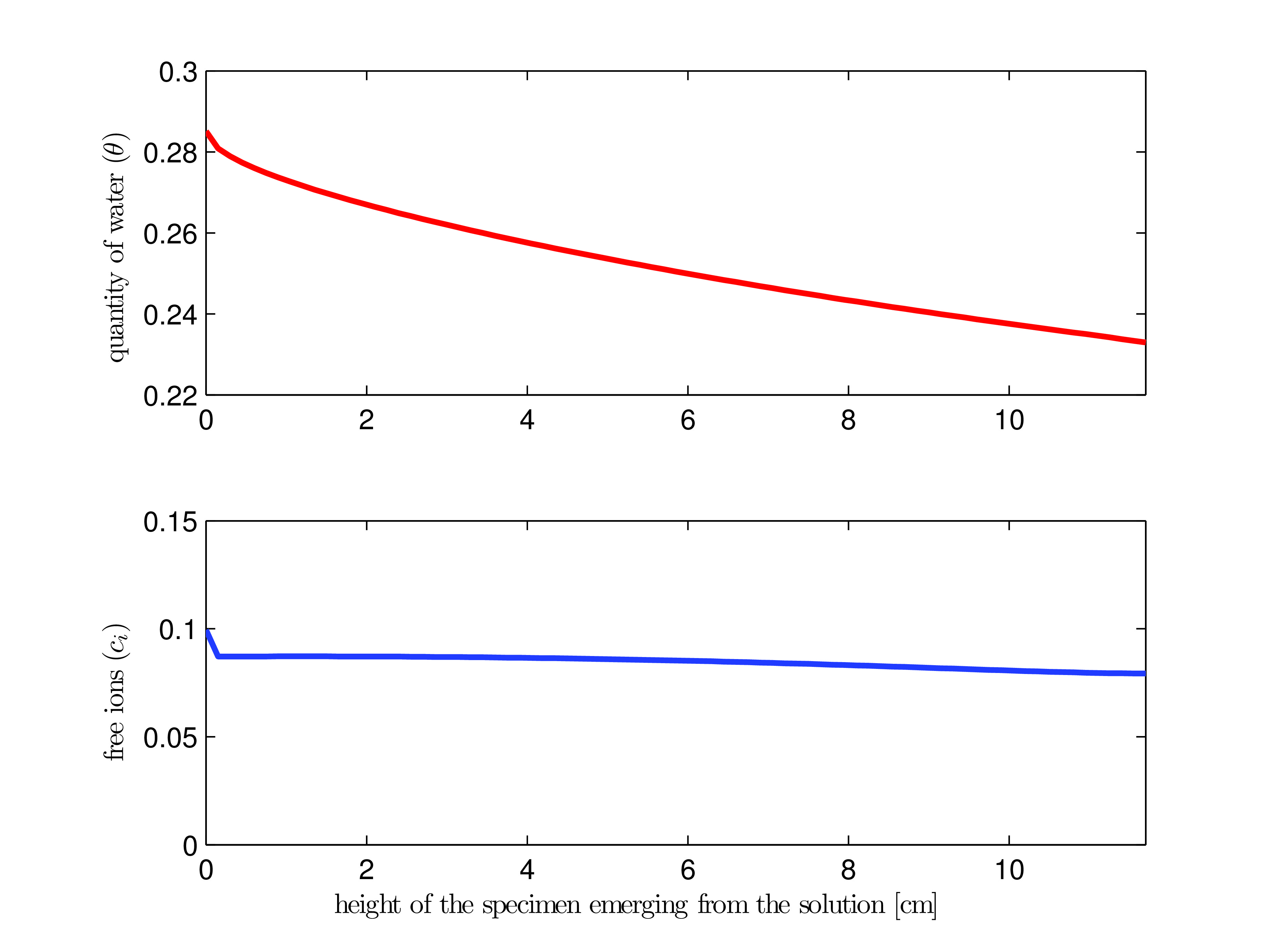}
\caption{Experiment 2. Imbibition phase in the salty solution: profile of $\theta_l$ and $c_i$ depicted at the final time of the experiment $T= 1128\ h$, with $K_s= 4.1 \cdot 10^{-5}\text{ s}^{-1}$ and $\gamma=0.6$.}
\label{fig:thetacs_imb}
\end{center}
\end{figure}

\begin{figure} [htbp!]
\begin{center}
\includegraphics[width=8cm,height=8cm]{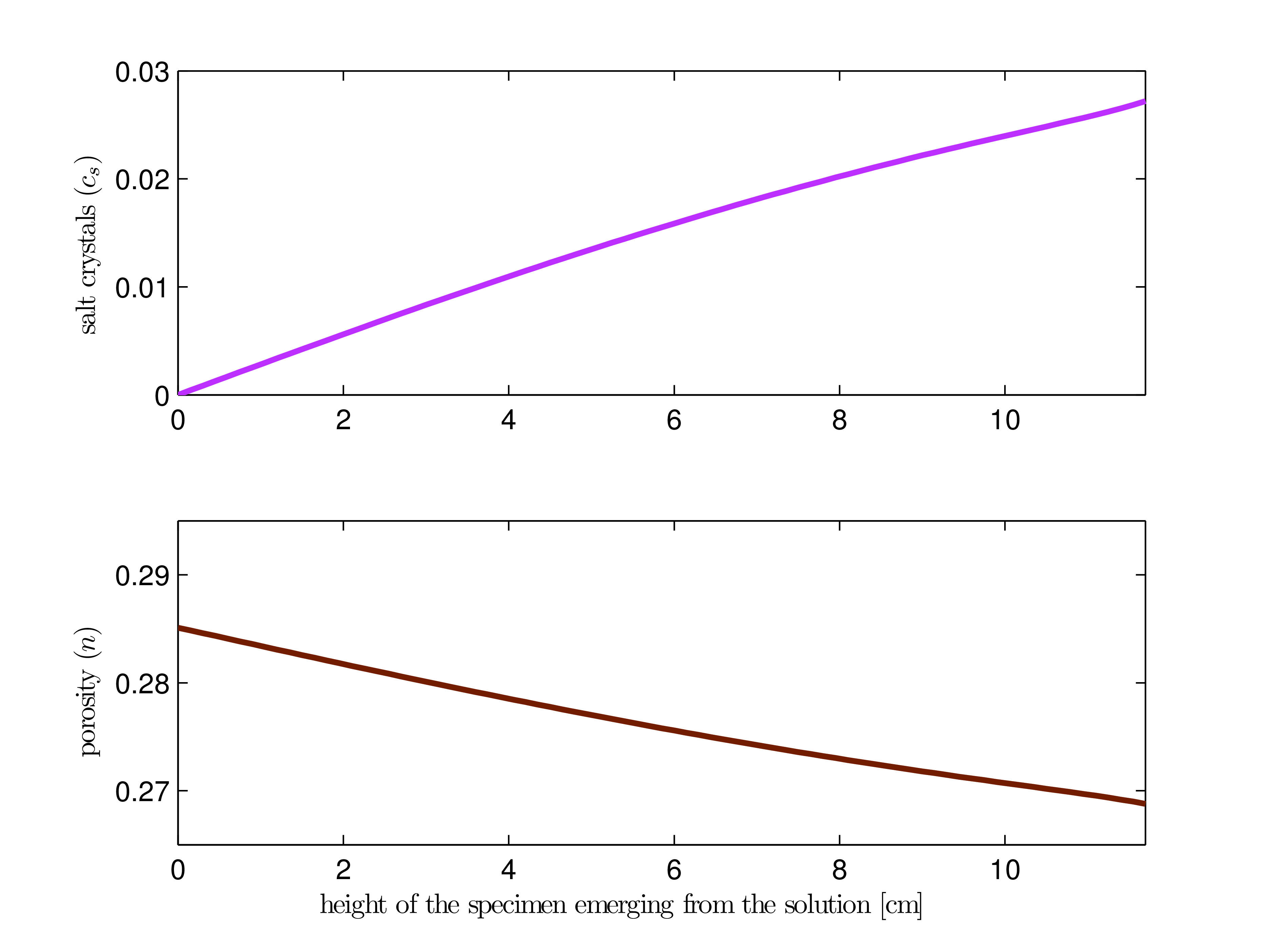}
\caption{Experiment 2. Imbibition phase in the salty solution: profile of $c_s$ and $n$ depicted at the final time of the experiment $T= 1128\ h$, with $K_s= 4.1 \cdot 10^{-5}\text{ s}^{-1}$ and $\gamma=0.6$.}
\label{fig:csn_imb}
\end{center}
\end{figure}

\begin{figure} [htbp!]
\begin{center}
\includegraphics[width=8cm,height=8cm]{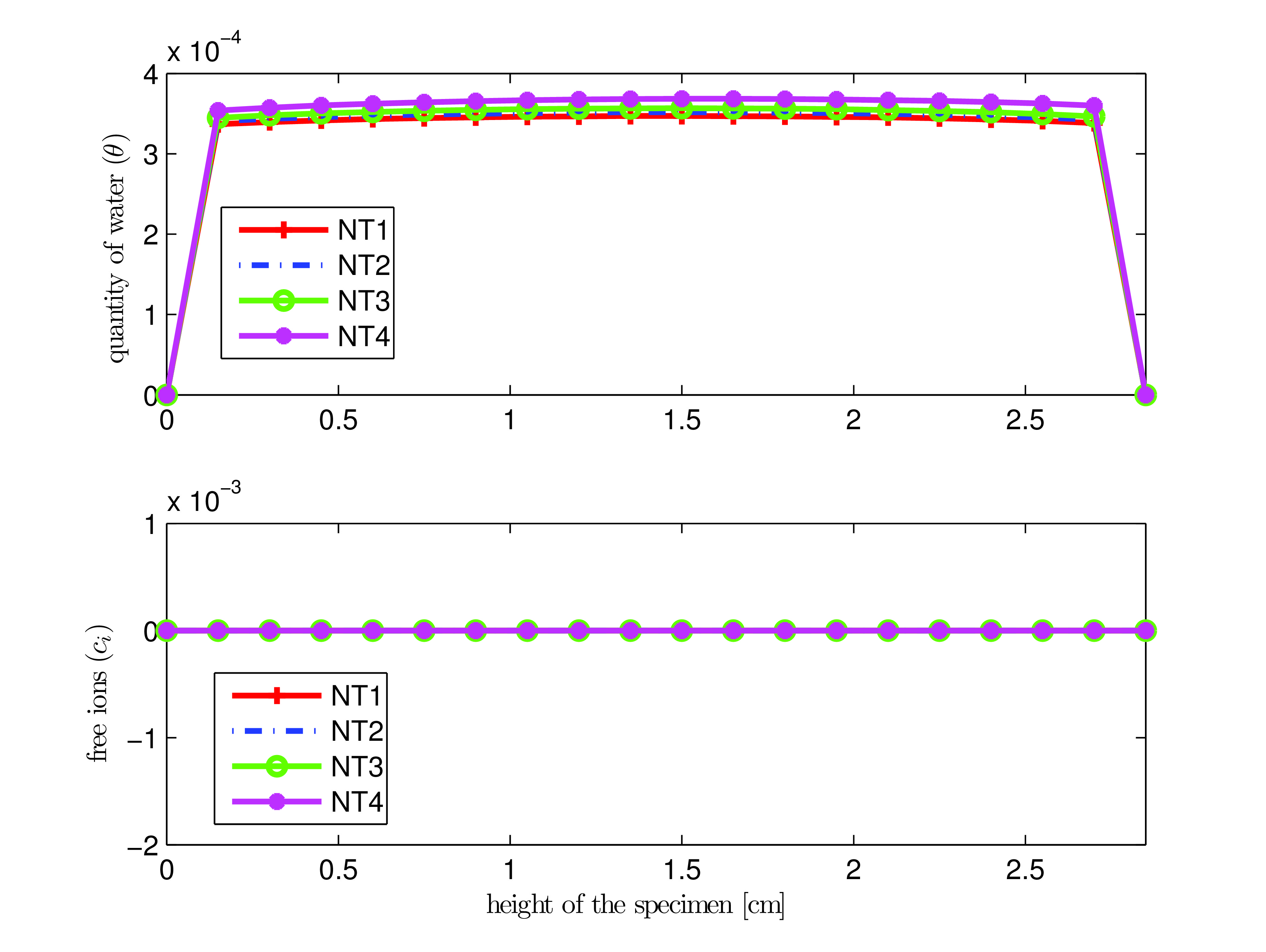}
\caption{Experiment 2. Drying phase: profile of $\theta_l$ and $c_i$, with $K_s= 4.1 \cdot 10^{-5}\text{ s}^{-1}$ and $\gamma=0.6$.}
\label{fig:thetacs_dry}
\end{center}
\end{figure}

\begin{figure} [htbp!]
\begin{center}
\includegraphics[width=8cm,height=8cm]{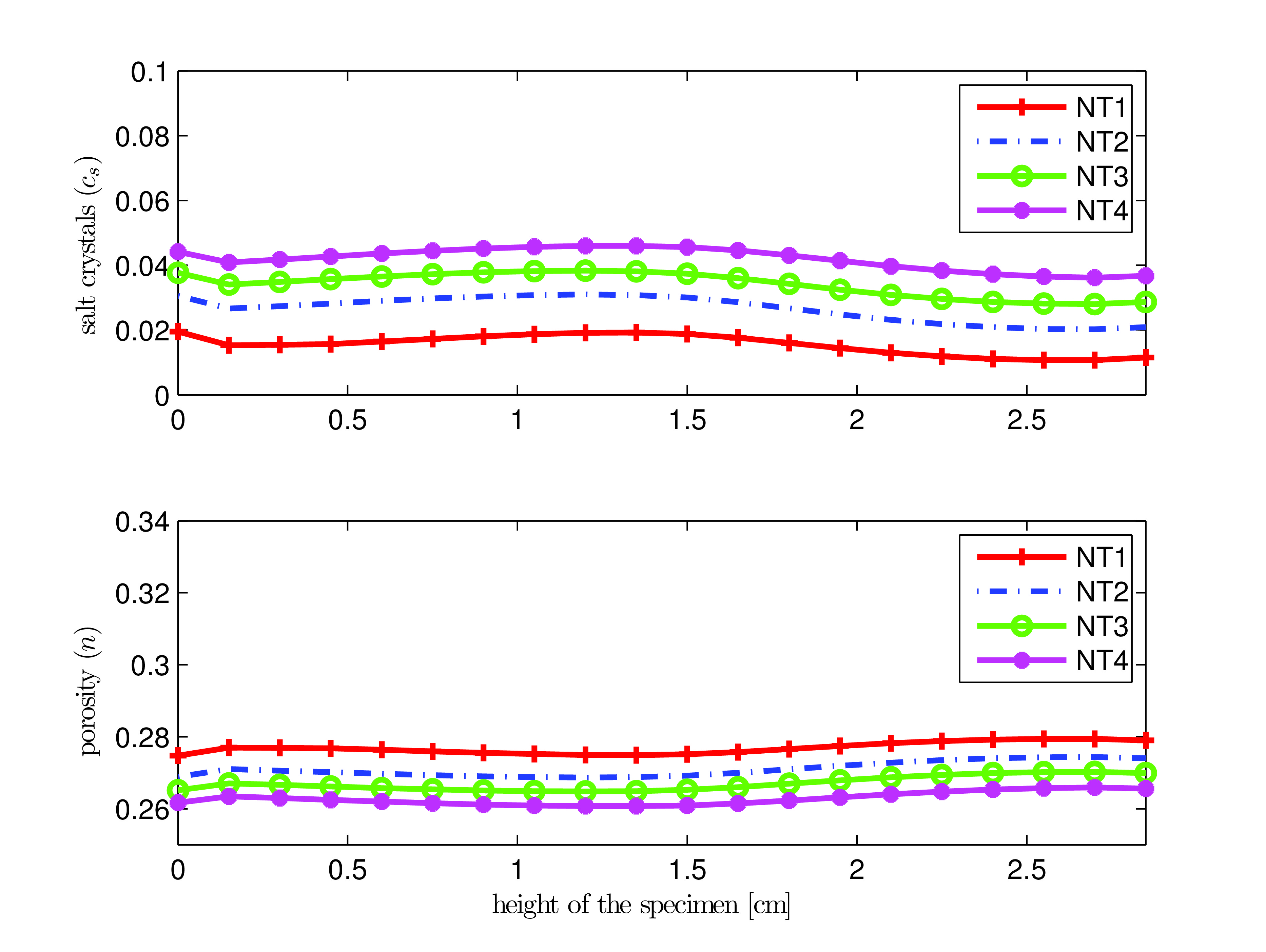}
\caption{Experiment 2. Drying phase: profile of $c_s$ and $n$, with $K_s= 4.1 \cdot 10^{-5}\text{ s}^{-1}$ and $\gamma=0.6$.}
\label{fig:csn_dry}
\end{center}
\end{figure}

For the case without inhibitor, the profiles of the quantities obtained numerically at the end of the imbibition experiment (47 days) for the not treated bricks, indicated by NT$i$, $i=1,2,3,4$, are depicted in Figg. \ref{fig:thetacs_imb} and \ref{fig:csn_imb}. As expected, the quantity of water in the brick $\theta_l$ is a decreasing function of the height of the brick, since the top of the brick is interested by water exchange with the exterior. The graphs of the same quantitites at the end of the drying phase are depicted in Figg. \ref{fig:thetacs_dry} and \ref{fig:csn_dry}. We observe that the amount of bound salts is, as expected, an increasing function of the height of the brick, since crystals mostly form where the quantity of water is lower. The calibration procedure gives the following result: we obtain an error of about $11.6\%$ for the values $K_s= 4.1 \cdot 10^{-5}\text{ s}^{-1}$ and $\gamma=0.6\text{ cm}^3 {g}^{-1}$.  In Table \ref{table:4} we report the comparison between measured data and numerical values obtained using the parameters deriving from the calibration procedure.

\begin{table}[h!]
\begin{center}
\begin{tabular}{|c|c|c|c|c|}
\hline
quantity & NT1 & NT2 & NT3 & NT4 \\
\hline
$q_i$&14.62&17.18&17.74&30.18\\
\hline
$q^{num}_i$& 12.21 & 17.88  & 22.52 &30.10\\
\hline
\end{tabular}
\end{center}
\caption{Salt content in any small brick in the not treated case (NT). We reported the measured salt content $q_i$ and the numerical values $q^{num}_i$ expressed in $mg/cm^3$.}
\label{table:4}
\end{table}

\begin{figure} [htbp!]
\begin{center}
\includegraphics[width=8cm,height=8cm]{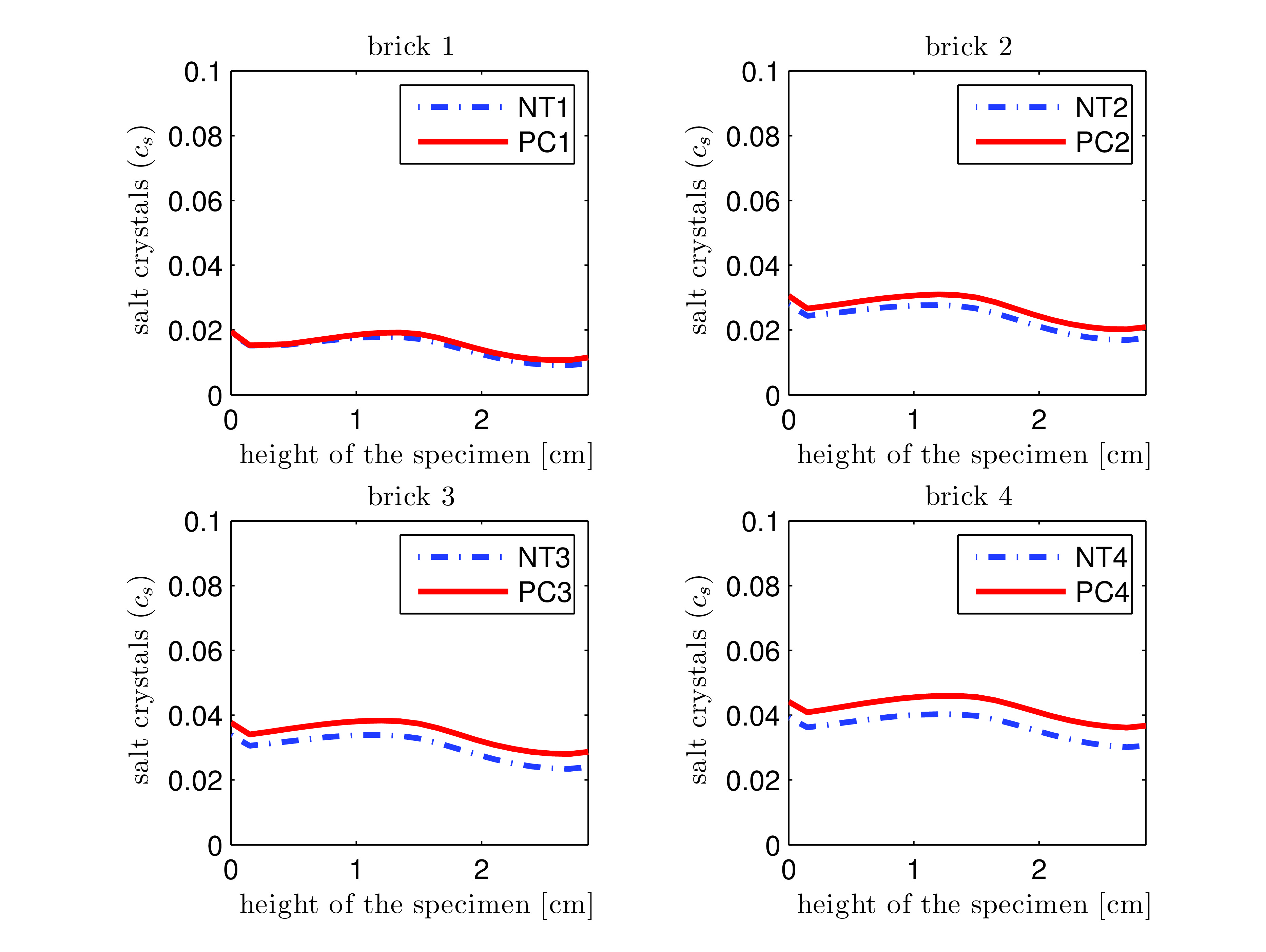}
\caption{Experiment 2. Comparison between the profile of $c_s$ without and in presence of PC-$10^{-6} M$.}
\label{fig:comparison}
\end{center}
\end{figure}

For the experiment of the bricks treated with PC-$10^{-6} M$, at the end of the calibration procedure we obtain an error of about $13.7\%$ for the values $ K_s= 6 \cdot 10^{-5}\text{ s}^{-1}$ and $\gamma=0.53\text{ cm}^{3}\text{ g}^{-1}$. In Table \ref{table:5} we report the comparison between measured data and numerical values obtained using the parameters deriving from the calibration procedure for the four bricks, indicated by PC$i$, $i=1,2,3,4$. In Fig. \ref{fig:comparison} we depicted the profile of $c_s$ for the not treated case (NT) and in presence of PC-$10^{-6} M$ (PC). As observed experimentally, the amount of salt crystals is higher in the case of the treatment with the crystallization modifier.

The computational time for a single simulation with fixed parameters both for the treated and the not-treated case takes 2240 seconds on an Intel(R) Core(TM) i7-3630QM CPU 2.4 GHz.

\begin{table} [h!]
\begin{center}
\begin{tabular}{|c|c|c|c|c|}
\hline
quantity & PC1 & PC2 & PC3 & PC4 \\
\hline
$q_i$&18.14 & 19.51  & 20.64 & 35.39\\
\hline
$q^{num}_i$&  13.53 & 21.40 & 26.78 & 35.56  \\
\hline
\end{tabular}
\end{center}

\caption{Salt content in any small brick in the treated case (PC). We reported the salt content $q_i$ and the numerical values $q^{num}_i$ expressed in $mg/cm^3$.}
\label{table:5}
\end{table}

Let us define the average porosity in $i$-th brick as:
\begin{equation}\label{por_num}
\frac{1}{h^b_i}\int_{a_{i}}^{b_i} n(z,\bar{t}) dz\text{ for }i=1,2,3,4,
\end{equation}
then for completeness, we report in the next Table \ref{table:6} the average porosity obtained numerically $n^{num}_i, \ i=1,\ldots,4$ for the four bricks obtained discretizing (\ref{por_num}) with the trapezoidal quadrature rule, both in the not treated (NT) and treated (PC) case.

\begin{table} [h!]
\begin{center}
\begin{tabular}{|c|c|c|c|c|}
\hline
porosity & $n^{num}_1$ &  $n^{num}_2$ &  $n^{num}_3$ &  $n^{num}_4$ \\
\hline
NT& 0.2355 & 0.2324 & 0.2280 & 0.2251\\
\hline
PC &  0.2331 & 0.2343 & 0.2264 & 0.2233  \\
\hline
\end{tabular}
\end{center}

\caption{Porosity in any small brick. We reported the average value for the porosity $n^{num}_i, i=1,\ldots,4$ for the not treated and the treated bricks.}
\label{table:6}
\end{table}

\section{Conclusions}
We developed a mathematical model to describe the action of crystallization inhibitors into a porous stone. This simple model is able to capture the main features of the inhibitor from experiments carried out on 
a set of commercially available bricks. According to the current knowledge, the model describes the action of inhibitors through two coefficients: crystallization rate, $K_s$, taking into account nucleation, and the specific volume $\gamma$, taking into account the crystal habit modification.
From the calibration of the mathematical model described in section \ref{sec:mathmodel}, we found out that the action of phosphocitrate (PC) increases the crystallization rate and decreases the crystal specific volume. This means that, although crystals form faster in the presence of the inhibitor, nevertheless they occupy a smaller volume, thus lowering the development of tensile stresses, and, on the other hand, ensuring the hydraulic continuity into the porous stones.
In the future, we will repeat the same study varying the materials and with more detailed experiments in order to test and improve our mathematical model. Our aim is to end up with a sound simulation tool to investigate crystallization modifier. 


\begin{description}
	\item[Acknowledgement] 
\end{description}
A. M. and C. R. gratefully acknowledge Regione Umbria (Italy) for funding through POR Umbria FSE 2007-2013, Asse IV "Capitale Umano" initiative.

\bibliographystyle{plain}

\end{document}